# Direct observation of quantum vortex fractionalization in multiband superconductors

Yu Zheng[1,*], Quanxin Hu[1,*,†], Haijiao Ji[1], Igor Timoshuk[2,3], Hanxiang Xu[4], Yongwei Li[1], Ye Gao[5], Xin Yu[4], Rui Wu[4], Xingye Lu[6], Vadim Grinenko[1,5], Egor Babaev[2,3], Noah F. Q. Yuan[1], Baiqing Lv[1,5,7,†], Chi-Ming Yim[1,5,†], Hong Ding[1,8,9]

[1] Tsung-Dao Lee Institute, Shanghai Jiao Tong University, Shanghai 201210, China

[2] Department of Physics, The Royal Institute of Technology, Stockholm SE-10691, Sweden

[3] Wallenberg Initiative Materials Science for Sustainability, Department of Physics, The Royal Institute of Technology, Stockholm SE-10691, Sweden

[4] Beijing National Laboratory for Condensed Matter Physics, Institute of Physics, Chinese Academy of Sciences, Beijing 100190, China

[5] School of Physics and Astronomy, Shanghai Jiao Tong University, Shanghai 200240, China.

[6] Center for Advanced Quantum Studies, School of Physics and Astronomy, Beijing Normal University, Beijing 100875, China

[7] Zhangjiang Institute for Advanced Study, Shanghai Jiao Tong University, Shanghai 200240, China.

[8] Hefei National Laboratory, Hefei 230088, China

[9] New Cornerstone Science Laboratory, Shanghai 201210, China

*These authors contributed equally to this work.
†Corresponding author. Email: sjtu18810999766@sjtu.edu.cn; baiqing@sjtu.edu.cn; c.m.yim@sjtu.edu.cn

**Magnetic field is expelled from a superconductor, unless it forms quantum vortices, consisting of a core singularity with current circulating around it. The London's quantization condition implies that there is one core singularity per quantum of magnetic flux in single-component superconductors, while in multiband materials fractional vortices are possible. Here, we report the first observation of quantum vortex core fractionalization on the potassium terminated surface of multiband superconductor $KFe_2As_2$ by scanning tunneling microscopy. We observe splitting of an integer-flux vortex into several fractional vortices, leading to disparity between numbers of flux quanta and vortex cores. Our findings demonstrate that fractionalized core singularities are possible in a multiband superconductor, opening avenue for new experimental platforms with quasiparticles with fractional statistics.**

The fundamental characteristic of superconductors is their ability to expel magnetic fields, a phenomenon known as the Meissner effect (*1*). However, magnetic field can penetrate superconductors if they form topological defects: quantum vortices. In type-II superconductors, purely repulsive vortex–vortex interaction leads to the formation of the Abrikosov lattice at magnetic fields above the first critical field (*2*). These quantum vortices manifest as line defects in three dimensions (3D) or point defects in two dimensions (2D), where the phase of the superconducting order parameter undergoes a winding of $2\pi$ or its integer multiple. Consequently, each quantum vortex carries a quantized magnetic flux of $\Phi_0 = h/2e$ (Fig. 1**A**) (*3*, *4*). The phase winding, responsible for flux quantization produces also a singularity: the so-called vortex core (*5*). This singularity is robust and mobile, hence a vortex is often referred as a particle-like topological soliton. Adding to particle-vortex duality, the lattice of singularities can melt forming vortex liquid. Given that, the quantization was considered as a fundamental principle of quantum fluids, the inquiry of possible existence of topological defects carrying a fractional portion of the flux quantum attracted significant research interests. Stationary objects carrying half of magnetic flux quantum were observed at grain boundaries of *d*-wave superconductors (*6–8*), small rings of $Sr_2RuO_4$ (*9*), and β-$Bi_2Pd$ (*10*) that were interpreted as a suggestion of possible existence of half-quantum vortices in these materials and hence unconventional pairing symmetries. However, numerous attempts to create and observe an actual half-quantum vortex, a mobile object with a core singularity in such systems, were not successful, despite significant motivation due to their potential applicability in topological quantum computing (*11–13*).

From the viewpoint of standard Ginzburg-Landau theory, the order parameter phase and its winding account for collective description of the entire electronic system, irrespective of microscopic detail, such as composition of band structure. The discovery of multiband superconductors has significantly enriched the field of superconductivity research. Due to the multiband nature of these materials, it is possible for electrons or holes in each Fermi surface to form a superconducting condensate, with interactions occurring between them as a result of electron or hole hopping between different bands. In multiband superconductors, vorticity (and consequently core singularity) can be supported in only one band. This introduces multiple phase fields in different bands, allowing for a phase that can be attributed to and have a winding in an individual band. Consequently, a vortex in such a system becomes "unquantized", meaning it can carry an arbitrary fraction of magnetic flux (*14*). The property of carrying an arbitrary fraction of the flux quantum and electrons in their cores allows these objects to be described using fractional statistics arguments (*15*), thus identifying them as anyons.

However, creation of fractional vortices in superconductors is nontrivial. Vortices in different components are attracted to each other due to their coupling to the same vector potential field and direct interband coupling. Consequently, in the ground state, fractional vortices in different condensates coalesce to form a "composite vortex" characterized by a single core singularity with the standard integer quantum flux $\Phi_0$. Recently, mobile vortices carrying a fraction of magnetic flux quantum were reported in magnetometry experiments on hole-doped $Ba_{1-x}K_xFe_2As_2$ (x ≈ 0.77) (*16*). However, despite of numerous spectroscopic studies of multiband superconductors, the fractionalization of vortex core singularities has not been observed. Here, for the first time, we demonstrate the splitting of single core singularities into fractional singularities on the potassium-terminated surface of the iron arsenide superconductor $KFe_2As_2$. We associate this dissociation phenomenon to the $s \pm is$ superconducting state with broken time-reversal symmetry (BTRS) (*17, 18*) and provide the spectroscopic characteristics of these fractional vortices.

**Topographic characterization of K-terminated surface**

$KFe_2As_2$, as a stoichiometric end member within the hole-doped 122-type iron arsenides, exhibits a tetragonal crystalline structure ($I4/mmm$), where layers of $Fe_2As_2$ are interleaved by layers of potassium (K) atoms. These K atoms serve as electron donors, furnishing an equal contribution of electrons to the neighboring $Fe_2As_2$ layers (Fig. 1**B**). The K- and arsenic (As)-terminated surfaces are predisposed to exposure owing to the weaker interlayer bonding between K and As layers. We focus our study on the K-terminated surface, as variation in the quantity of retained K atoms on the cleavage plane may impart unique properties that differ from those observed in the bulk states. Two types of ordered surface structures can be distinguished for the K-terminated surface (Fig. 1**C**). An atomically resolved topographic image of the K-terminated surface (Fig. 1**D**) taken from the blue region in Fig. 1**C** shows the most commonly observed reconstruction (*19*). The observed lattice constants exhibit a $\sqrt{2} \times \sqrt{2}$ expansion relative to the uncleaved K plane, with lattice directions aligning along the Fe-Fe directions, as indicated by the white arrows. We designate this reconstruction as "rt2". The second termination in Fig. 1**C** (red region) shows an identical $1 \times 1$ lattice symmetry as the K layers in the bulk. On this basis, we designate this termination as the unreconstructed K- terminated surface, or as "$1 \times 1$" in the remaining context. Shown in Fig. 1**F**, the differential conductance d$I$/d$V$ spectrum taken from the "rt2" surface is more consistent with the bulk properties: there is a van

Hove singularity (vHs) peak at -7 meV, which stems from the saddle points located in the middle of the principal axes of the first Brillouin zone (*19*). In contrast, the tunneling spectrum from "1 × 1" surface differs significantly from that of the "rt2" surface. The characteristic peak observed at +66 meV resembles the hump in the d$I$/d$V$ spectrum at +33 meV of LiFeAs, which was attributed to the band top of the outermost $d_{xy}$ orbital (*20*). Moreover, it is noteworthy that the two types of terminations in Fig. 1**C** indeed form from the same layer of K atoms. The apparent height difference (~ 3 Å) between the two terminations originates from the convolution between the integrated local density of states and spatial corrugations within the constant current topographic image (See Fig. S1 for more data and discussions).

The "rt2" reconstruction involves the removal of half of the K atoms in the region to maintain charge neutrality, thus preserving an equivalent doping level on the cleaved surface as compared to the bulk material. Conversely, the "1 × 1" termination exhibits an excess of 50% K atoms. Quasiparticle interference (QPI) measurements conducted on these two terminations reveal a downward shift of the hole band by ~10 meV, thus confirming the significant electron doping effects (See Fig. S2 for more data). This provides us with a stoichiometric platform to investigate 122-type iron arsenides in heavily hole-doped regime.

**Enhanced superconductivity and vortex bound states (VBS)**

The surface electron doping effect significantly impacts the material properties within the proximal layers of the surfaces, resulting in deviations from those of $KFe_2As_2$ and making them more reminiscent of heavily hole-doped 122-type iron arsenides. The large energy-scale d$I$/d$V$ spectrum measured on the "1 × 1" surface resembles that obtained from the Ba/K surface of $Ba_{1-x}K_xFe_2As_2$ (x ≈ 0.77), which attracts strong interest due to experimental observations of BTRS (*21–23*). Both display a d$I$/d$V$ peak at comparable positive energy values (Fig. 2**A**). According to the phase diagram of 122-type superconductor $Ba_{1-x}K_xFe_2As_2$ (*24*), electron doping of $KFe_2As_2$ is expected to enhance superconductivity. Unlike the spectrum measured on "rt2" surface (*19*), that of the "1 × 1" surface yields superconducting coherence peaks at ±3.1 meV (Fig. 2**B**) and depleted density of states (DOS) around the zero energy as a result of superconducting state formation. Temperature dependent d$I$/d$V$ measurements reveal that the superconducting gap closes at ~16 K (Fig. 2**B**), nearly five times of the superconducting transition temperature of $KFe_2As_2$ ($T_c$ ~ 3 K). Further density functional theory (DFT) calculations (Fig. S3) confirm that the doping level of the proximal layers of the "1 × 1" surface closely resembles that of $Ba_{1-x}K_xFe_2As_2$ (x ≈ 0.77), consistent with a simple counting for the top layer x = (2+1)/2/2 = 0.75.

In addition to enhanced superconductivity, the "1 × 1" surface also exhibits other distinct properties from the bulk materials. To demonstrate this, we performed zero-bias conductance (ZBC) map measurements on the "1 × 1" surface at a temperature of 300 mK with an external magnetic field of 3 T applied along the crystal c-axis, which exceeds the upper critical field of $KFe_2As_2$ (*25*). Like LiFeAs (*26*), the ZBC map around a vortex displays a star-like shape (Fig. 2**C**), but whose tails are along the Fe-Fe directions. We performed d$I$/d$V$ measurements along the Fe-Fe and As-As directions across a vortex core (Fig. 2**D**), and the results are shown in Figs. 2**E** to **H**. Along the As-As direction, the pronounced spectral feature is a pair of dispersive VBS at non-zero energies (Figs. 2**E** and **G**). Owing to the high resolution of our data, two additional dispersive VBS at non-zero energies can be distinguished along the Fe-Fe direction (Figs. 2**F** and **H**). The extracted energy positions of the VBS are shown as solid circles. Through analysis, we can attribute the outer dispersive VBS and inner dispersive VBS along the Fe-Fe direction to the inner $d_{xz}$ band and outer $d_{yz}$ band respectively (See SM for more discussion).

**Quantum vortex splitting and fractional vortices**

As discussed above, the doping level of the proximal layers of the "1 × 1" surface closely resembles that of $Ba_{1-x}K_xFe_2As_2$ (x ≈ 0.77). This provides an opportunity to study the quantum vortex fractionalization on this new two-dimensional stoichiometric platform in the vicinity of this magic doping. In multiband superconductor, if each superconducting component produces a nonoverlapping core singularity, electronic fingerprints indicative of quantum vortex fractionalization is expected on the "1 × 1" surface using STM.

To search for the spectroscopic evidence of fractional vortices, we measured temperature dependent ZBC maps at the same field of view with an external magnetic field of 2 T, and the results are shown in Fig. 3**A**. At 1.8 K, nearly all vortices maintain a star-like shape. As temperature increases (2.5-3 K), some still retain a star-like shape while others start to deform. Further increasing the temperature to 3.5 K, some ordinary vortices, marked by red dashed ellipses in Fig. 3**A**, undergo a splitting process, leading to extra signals clearly present in their vicinity. Further increasing the temperature to 4.2 K, it becomes even clearer that each of few ordinary vortices (marked by the red dashed ellipses) splits into two or three anomalous vortices, with their DOS at the vortex center significantly weaker than those of ordinary vortices. Figures 3**B** to **D** illustrate the temperature evolution of zoomed-in ZBC maps around vortex cores #N1, #F4, and #F6 in Fig. 3**A**. The spatial dependence of the ZBC intensity along the red arrow lines crossing the vortex cores (Fig. 3**E**) clearly illustrates the splitting of the magnetic flux vortices. The red dashed lines mark the positions of the vortex cores (See Fig. S4 for more splitting vortices). It is evident that some vortices, such as #N1, remain as isolated one-quanta vortices, while others, such as #F4 and #F6, undergo splitting with increasing temperature, forming multiple anomalous vortices.

Given the total magnetic flux $\Phi$ through the sample, one can calculate the expected number of vortices $N = \Phi/\Phi_0$ by assuming standard situation where a vortex carries a quantum of magnetic flux. As plotted in Fig. 3**F**, we performed a statistical comparison between the expected number of vortices and the number of experimentally observed vortices within a (365 × 375) nm$^2$ area, as a function of magnetic field strength (See Fig. S5 and Table S1 for more data), with the imaged area carefully corrected according to the locations of defects. At 300 mK (left part of Fig. 3**F**), the expected number of vortices closely matches with the number of experimentally observed vortices on the "1 × 1" surface (indicated by purple markers). However, at 4.2 K (right part of Fig. 3**F**), the number of experimentally observed vortices significantly exceeds the expected number. We note that a similar increase of vortex number has been noticed in a previous work (*27*, *28*). Furthermore, the number of normal vortices is lower than the expected number. The ratio of the number of anomalous vortices to the discrepancy between expected number and observed number of normal vortices falls within the range of 2 to 3, consistent with the evolutionary trend of vortex splitting. This indicates that the anomalous vortices observed here are fractional vortices formed from the splitting of normal vortices. In Fig. 3F, the right-hand axis displays the average magnetic flux associated with anomalous vortices. Under this condition, the anomalous vortex exhibits a fractional magnetic flux. Our observation of the coexistence of various types of vortex cores aligns with previous findings of coexisting integer-flux and fractional-flux vortices in superconducting quantum interference device (SQUID) magnetometry studies of hole-doped $Ba_{1-x}K_xFe_2As_2$ (x ≈ 0.77) (*16*).

Another alternative explanation for anomalous number of vortex cores without violation of flux quantization would be spontaneous formation of vortex-antivortex pairs. Hence it is important to obtain direct spectroscopic evidence for the presence of fractional vortices. We obtain that evidence bas in the fact that at the core of a fractional vortex, some components of the superconducting order parameters vanish, while the other components can remain approximately the same as those away from the core (*14*). Consequently, compared to normal vortices, in fractional vortices, VBS intensity is expected to be weaker and superconducting coherence peaks are stronger. Figure 3**G** shows the line-cut d$I$/d$V$ intensity plots across the normal vortex #N2 and anomalous vortex #F6 at 4.2 K. The VBS of the normal vortex at 4.2 K resemble those at low temperatures (Fig. 2**D**), albeit with decreased energy resolution as a result of temperature increase. However, the anomalous vortex exhibits distinct characteristics, notably a reduction in VBS intensity at the vortex core. Figure 3**H** shows the d$I$/d$V$ spectra measured at the core centers of multiple normal and anomalous vortices. The VBS intensities measured at the cores of normal vortices exceed those of anomalous vortices. Furthermore, the superconducting coherence peaks at the cores of normal vortices are smaller than those of anomalous vortices (See Fig. S6 for more data).

**Discussion**

To see how a fractional vortex is different from integer vortex, we preform theoretical simulations of the spectroscopic characteristics of both the normal and fractional (or splitted) vortices, with the corresponding calculated DOS at the zero energy as a function of position maps displayed in Figs. 4**A** and **B** respectively. In Fig. 4**C**, the calculated DOS linecuts along the As-As direction before and after vortex splitting are consistent with the experimental results shown in Fig. 3**G**. Compared to fractional vortices, the calculated DOS spectra at the cores of normal vortices exhibit stronger VBS intensity but weaker superconducting coherence peaks (Fig. 4**D**). The predicted spectroscopic characteristics of fractional vortices are therefore consistent with the experimental results. Additionally, the self-consistent solutions of the Bogoliubov-de Gennes model also qualitatively agree with the experimental observations (Figs. S8**A-C**). The model, methodology, and details of the simulation are provided in the Supplementary Material (SM).

We emphasize that in standard models of multiband superconductors, although the phase winding numbers for superconducting condensates in different bands may differ, potentially resulting in vortices carrying fractional quantum flux, nonetheless, a bound state as a "composite vortex" characterized by a single core singularity is energetically favored in a large system due to generic intercomponent coupling by vector potential (*14*) and interband Josephson coupling. As with other typical multiband iron-based superconductors, we did not observe any similar phenomenon of vortex splitting on LiFeAs with $s_\pm$ pairing states (indicated by goldenrod markers in Fig. 3**F** and Fig. S5I). The surface electron doping effect in $KFe_2As_2$ makes the proximal layers of the surfaces more similar to $Ba_{1-x}K_xFe_2As_2$ (x ≈ 0.77). The $s \pm is$ superconducting state with spontaneous BTRS has garnered theoretical (*29, 30*) and robust experimental (*21–23*) support in the vicinity of $Ba_{0.23}K_{0.77}Fe_2As_2$. This insight prompts us to consider the possibility that time-reversal symmetry might be broken on the "1 × 1" surface. These BTRS states break the $U(1) \times Z_2$ symmetry. Domain wall excitations associated with the breakdown of discrete $Z_2$ symmetry segregate regions of different broken states. Pinned domain walls can help formation of fractional vortices: when an integer flux vortex in an *N*-band superconductor is placed on the domain wall, Josephson interactions energetically favor splitting an integer flux vortex into *N* fractional vortices, each with independent spatially separated singularities in *N* bands. Such defects can form aggerates characterized by a new topological index and called to as "chiral $CP^{N-1}$ skyrmions" (*17, 18, 31*). We observe that some integer vortices split into three fractional vortices. Hence the number of fractional vortices in this system is larger than the number of broken symmetries, hence indicating that the fractionalization is likely associated with the phenomenon of vorticity in different bands rather the different phenomenon of formation of half-quantum vortices possible in p-wave superconductors (*32*).

In addition to the anomalous vortices observed in the ZBC maps, as illustrated in Fig. S7, distinct chain features are prominently discernible. We highlight these features in Fig. S7 for visual reference. The chain features exhibit a higher DOS at zero bias. Moreover, the anomalous vortices align along these chain features. This observation is consistent with the occurrence of $CP^{N-1}$ skyrmions in the form of fractional vortices along domain walls. Self-consistent solutions of the Bogoliubov-de Gennes model is consistent with the observation that the domain wall being uniformly decorated by fractional vortices in an external field (see more details in Fig. S8). The emergence of these chain features suggests the possible presence of BTRS superconducting states. We also note that the two dimensionality of this $s \pm is$ superconducting surface, promoting fluctuation-induced closed domain walls, might facilitate the formation of fractional vortices.

In Fig. 4**E**, we present the phase diagram of the "1 × 1" surface. At low temperatures, although the "1 × 1" surface exhibits BTRS superconductivity with limited number of domain walls, the formation of fractional vortices suppressed. Most vortices persist in the form of integer flux vortices. As temperature gradually increases, thermal fluctuations provide the necessary entropy leading to the splitting of integer flux vortices, particularly those located at domain walls, into fractional vortices. When temperature exceeds $T_c^{VS} \approx 3$ K, the fractionalization of vortex core singularities has all been stabilized and formed.

Our work reports the first observation of fractionalization of the core of the quantum vortex in a multiband superconductor, along with spectroscopic characteristics of the resulting fractional vortices. We argue that time-reversal symmetry might be broken on the "1×1" surface of $KFe_2As_2$, which together with the two dimensionality facilitates the splitting of the vortex core singularity. Crucially, we provide a general methodology for revealing fractional vortices in superconductors using STM. The fractional vortices might be resolved in some other iron-based superconductors, like $Ba(Fe_{1-x}Co_x)_2As_2$ (*33, 34*) and $BaFe_{2-x}Ni_xAs_2$ (*35*) by revisiting STM measurement with super-high spatial resolution. The STM-based observation of fractional vortices is especially important because these objects can obey fractional statistics by realizing electron-fractional-flux-composite (*15*). Our findings not only provide a new two-dimensional stoichiometric platform for studying fractional vortices with mobile core singularity, that can be utilized as a new fluxonics platform (*36, 37*), but also will stimulate intensive exploration of the multiband nature and superconductivity with broken time-reversal symmetry in iron-based superconductors.

**References**

1. W. Meissner, R. Ochsenfeld, Ein neuer Effekt bei Eintritt der Supraleitfähigkeit. *Naturwissenschaften* **21**, 787–788 (1933).
2. AA Abrikosov, On the magnetic properties of superconductors of the second group. *Soviet Physics-JETP* **5**, 1174–1182.
3. L. Onsager, Magnetic Flux Through a Superconducting Ring. *Phys. Rev. Lett.* **7**, 50–50 (1961).
4. F. London, On the Problem of the Molecular Theory of Superconductivity. *Phys. Rev.* **74**, 562–573 (1948).
5. L. Onsager, Statistical hydrodynamics. *Il Nuovo Cimento (1943-1954)* **6**, 279–287 (1949).
6. J. R. Kirtley, C. C. Tsuei, M. Rupp, J. Z. Sun, L. S. Yu-Jahnes, A. Gupta, M. B. Ketchen, K. A. Moler, M. Bhushan, Direct Imaging of Integer and Half-Integer Josephson Vortices in High- Tc Grain Boundaries. *Phys. Rev. Lett.* **76**, 1336–1339 (1996).
7. R. G. Mints, I. Papiashvili, J. R. Kirtley, H. Hilgenkamp, G. Hammerl, J. Mannhart, Observation of Splintered Josephson Vortices at Grain Boundaries in YBa$_2$Cu$_3$O$_{7-\delta}$. *Phys. Rev. Lett.* **89**, 067004 (2002).
8. H. Hilgenkamp, Ariando, H.-J. H. Smilde, D. H. A. Blank, G. Rijnders, H. Rogalla, J. R. Kirtley, C. C. Tsuei, Ordering and manipulation of the magnetic moments in large-scale superconducting π-loop arrays. *Nature* **422**, 50–53 (2003).
9. J. Jang, D. G. Ferguson, V. Vakaryuk, R. Budakian, S. B. Chung, P. M. Goldbart, Y. Maeno, Observation of Half-Height Magnetization Steps in Sr$_2$RuO$_4$. *Science* **331**, 186–188 (2011).
10. Y. Li, X. Xu, M.-H. Lee, M.-W. Chu, C. L. Chien, Observation of half-quantum flux in the unconventional superconductor β-Bi$_2$Pd. *Science* **366**, 238–241 (2019).
11. N. Read, D. Green, Paired states of fermions in two dimensions with breaking of parity and time-reversal symmetries and the fractional quantum Hall effect. *Phys. Rev. B* **61**, 10267–10297 (2000).
12. D. A. Ivanov, Non-Abelian Statistics of Half-Quantum Vortices in *p*-Wave Superconductors. *Phys. Rev. Lett.* **86**, 268–271 (2001).
13. C. Nayak, S. H. Simon, A. Stern, M. Freedman, S. Das Sarma, Non-Abelian anyons and topological quantum computation. *Rev. Mod. Phys.* **80**, 1083–1159 (2008).
14. E. Babaev, Vortices with Fractional Flux in Two-Gap Superconductors and in Extended Faddeev Model. *Phys. Rev. Lett.* **89**, 067001 (2002).
15. F. Wilczek, Magnetic Flux, Angular Momentum, and Statistics. *Phys. Rev. Lett.* **48**, 1144–1146 (1982).
16. Y. Iguchi, R. A. Shi, K. Kihou, C.-H. Lee, M. Barkman, A. L. Benfenati, V. Grinenko, E. Babaev, K. A. Moler, Superconducting vortices carrying a temperature-dependent fraction of the flux quantum. *Science* **380**, 1244–1247 (2023).
17. J. Garaud, J. Carlström, E. Babaev, Topological Solitons in Three-Band Superconductors with Broken Time Reversal Symmetry. *Phys. Rev. Lett.* **107**, 197001 (2011).
18. J. Garaud, J. Carlström, E. Babaev, M. Speight, Chiral *CP*$^2$ skyrmions in three-

band superconductors. *Phys. Rev. B* **87**, 014507 (2013).

19. D. Fang, X. Shi, Z. Du, P. Richard, H. Yang, X. X. Wu, P. Zhang, T. Qian, X. Ding, Z. Wang, T. K. Kim, M. Hoesch, A. Wang, X. Chen, J. Hu, H. Ding, H.-H. Wen, Observation of a Van Hove singularity and implication for strong-coupling induced Cooper pairing in $KFe_2As_2$. *Phys. Rev. B* **92**, 144513 (2015).

20. L. Kong, L. Cao, S. Zhu, M. Papaj, G. Dai, G. Li, P. Fan, W. Liu, F. Yang, X. Wang, S. Du, C. Jin, L. Fu, H.-J. Gao, H. Ding, Majorana zero modes in impurity-assisted vortex of LiFeAs superconductor. *Nat Commun* **12**, 4146 (2021).

21. V. Grinenko, R. Sarkar, K. Kihou, C. H. Lee, I. Morozov, S. Aswartham, B. Büchner, P. Chekhonin, W. Skrotzki, K. Nenkov, R. Hühne, K. Nielsch, S.-L. Drechsler, V. L. Vadimov, M. A. Silaev, P. A. Volkov, I. Eremin, H. Luetkens, H.-H. Klauss, Superconductivity with broken time-reversal symmetry inside a superconducting s-wave state. *Nat. Phys.* **16**, 789–794 (2020).

22. V. Grinenko, D. Weston, F. Caglieris, C. Wuttke, C. Hess, T. Gottschall, I. Maccari, D. Gorbunov, S. Zherlitsyn, J. Wosnitza, A. Rydh, K. Kihou, C.-H. Lee, R. Sarkar, S. Dengre, J. Garaud, A. Charnukha, R. Hühne, K. Nielsch, B. Büchner, H.-H. Klauss, E. Babaev, State with spontaneously broken time-reversal symmetry above the superconducting phase transition. *Nat. Phys.* **17**, 1254–1259 (2021).

23. I. Shipulin, N. Stegani, I. Maccari, K. Kihou, C.-H. Lee, Q. Hu, Y. Zheng, F. Yang, Y. Li, C.-M. Yim, R. Hühne, H.-H. Klauss, M. Putti, F. Caglieris, E. Babaev, V. Grinenko, Calorimetric evidence for two phase transitions in $Ba_{1-x}K_xFe_2As_2$ with fermion pairing and quadrupling states. *Nat Commun* **14**, 6734 (2023).

24. T. Shibauchi, A. Carrington, Y. Matsuda, Quantum critical point lying beneath the superconducting dome in iron-pnictides. *Annu. Rev. Condens. Matter Phys.* **5**, 113–135 (2014).

25. M. Abdel-Hafiez, S. Aswartham, S. Wurmehl, V. Grinenko, C. Hess, S.-L. Drechsler, S. Johnston, A. U. B. Wolter, B. Büchner, H. Rosner, L. Boeri, Specific heat and upper critical fields in $KFe_2As_2$ single crystals. *Phys. Rev. B* **85**, 134533 (2012).

26. T. Hanaguri, K. Kitagawa, K. Matsubayashi, Y. Mazaki, Y. Uwatoko, H. Takagi, Scanning tunneling microscopy/spectroscopy of vortices in LiFeAs. *Phys. Rev. B* **85**, 214505 (2012).

27. Liu L., "Scanning Tunneling Microscopy/Spectroscopy Study of Surface Enhanced Superconductivity of $KFe_2As_2$ and Charge-Density Wave of $ZrTe_3$," thesis, University of Chinese Academy of Sciences (Institute of Physics, Chinese Academy of Sciences) (2022).

28. Zhu C., "STM Study in Superconductor of $NbC/TaC/ZrTe_3/KFe_2As_2$," thesis, University of Chinese Academy of Sciences (Institute of Physics, Chinese Academy of Sciences) (2022).

29. S. Maiti, A. V. Chubukov, *s*+i*s* state with broken time-reversal symmetry in Fe-based superconductors. *Phys. Rev. B* **87**, 144511 (2013).

30. J. Böker, P. A. Volkov, K. B. Efetov, I. Eremin, *s*+i*s* superconductivity with incipient bands: Doping dependence and STM signatures. *Phys. Rev. B* **96**, 014517 (2017).

31. A. Benfenati, M. Barkman, E. Babaev, Demonstration of $CP^2$ skyrmions in three-

band superconductors by self-consistent solutions of a Bogoliubov--de Gennes model. *Phys. Rev. B* **107**, *094503 (2023)*.

32. M. Sigrist, K. Ueda, Phenomenological theory of unconventional superconductivity. *Rev. Mod. Phys.* **63**, *239–311 (1991)*.

33. B. Kalisky, J. R. Kirtley, J. G. Analytis, J.-H. Chu, I. R. Fisher, K. A. Moler, Behavior of vortices near twin boundaries in underdoped Ba(Fe$_{1-x}$Co$_x$)$_2$As$_2$. *Phys. Rev. B* **83**, *064511 (2011)*.

34. S. Ghosh, M. S. Ikeda, A. R. Chakraborty, T. Worasaran, F. Theuss, L. B. Peralta, P. M. Lozano, J.-W. Kim, P. J. Ryan, L. Ye, A. Kapitulnik, S. A. Kivelson, B. J. Ramshaw, R. M. Fernandes, I. R. Fisher, Elastocaloric evidence for a multicomponent superconductor stabilized within the nematic state in Ba(Fe$_{1-x}$Co$_x$)$_2$As$_2$. arXiv arXiv:2402.17945 [Preprint] (2024). https://doi.org/10.48550/arXiv.2402.17945.

35. L. J. Li, T. Nishio, Z. A. Xu, V. V. Moshchalkov, Low-field vortex patterns in the multiband BaFe2-xNixAs2 superconductor (x = 0.1, 0.16). *Phys. Rev. B* **83**, *224522 (2011)*.

36. T. Golod, A. Iovan, V. M. Krasnov, Single Abrikosov vortices as quantized information bits. *Nat Commun* **6**, *8628 (2015)*.

37. K. Miyahara, M. Mukaida, K. Hohkawa, Abrikosov vortex memory. *Applied Physics Letters* **47**, *754–756 (1985)*.

**Acknowledgments:**

We thank Fazhi Yang and Lingyuan Kong for useful discussions. We thank Ruidan Zhong for assistance in performing PPMS experiments. A portion of this work was carried out at the Synergetic Extreme Condition User Facility (SECUF).

**Funding:**

H.D. acknowledges support from the New Cornerstone Science Foundation (No. 23H010801236), Innovation Program for Quantum Science and Technology (No. 2021ZD0302700). B.L. acknowledges support from the Ministry of Science and Technology of China (2023YFA1407400), the National Natural Science Foundation of China (12374063), the Shanghai Natural Science Fund for Original Exploration Program (23ZR1479900), the TDLI starting up grant and the Shanghai talent Program. C.M.Y. acknowledges support from Shanghai Pujiang Talent Program (No. 21PJ1405400), TDLI Start-up Fund. Q.H. acknowledges support from China Postdoctoral Science Foundation (No. GZB20230421). E.B. acknowledges support by the Swedish Research Council Grants 2022-04763, by Olle Engkvists Stiftelse, and the Wallenberg Initiative Materials Science for Sustainability (WISE) funded by the Knut and Alice Wallenberg Foundation. V.G. acknowledges support from the National Natural Science Foundation of China (No. 1237040280). R.W. acknowledges support from the Strategic Priority Research Program of Chinese Academy of Sciences (No.

XDB33020300). H. J. and N. F. Q. Y. acknowledge support from the National Natural Science Foundation of China (No. 12174021).

**Author contributions:**

Q.H. H.D., C.M.Y. and B.L. conceived the experiment. Q.H., Y.Z. and Y.G. performed STM measurements and data analysis in the laboratory of C.M.Y.. Additional STM data were collected by Q.H. with the assistance of X.Y. and R.W. at the ultra-low temperature STM station (A7) of SECUF.  H.J., I.T., N.F.Q.Y. performed numerical simulations and theoretical analysis. E.B. performed theoretical analysis. H.X. performed DFT calculations. X.L. synthesized $KFe_2A_2$ single crystals. Y.L. and V.G. prepared and characterized (or selected) samples for the research. Q.H., Y.Z., B. L., C.M.Y., and H.D. wrote the paper with contributions from all authors. H.D., C.M.Y. and B.L. supervised the project.

**Competing interests:** Authors declare that they have no competing interests.

**Data and materials availability:** All data are available in the main text or the supplementary materials.

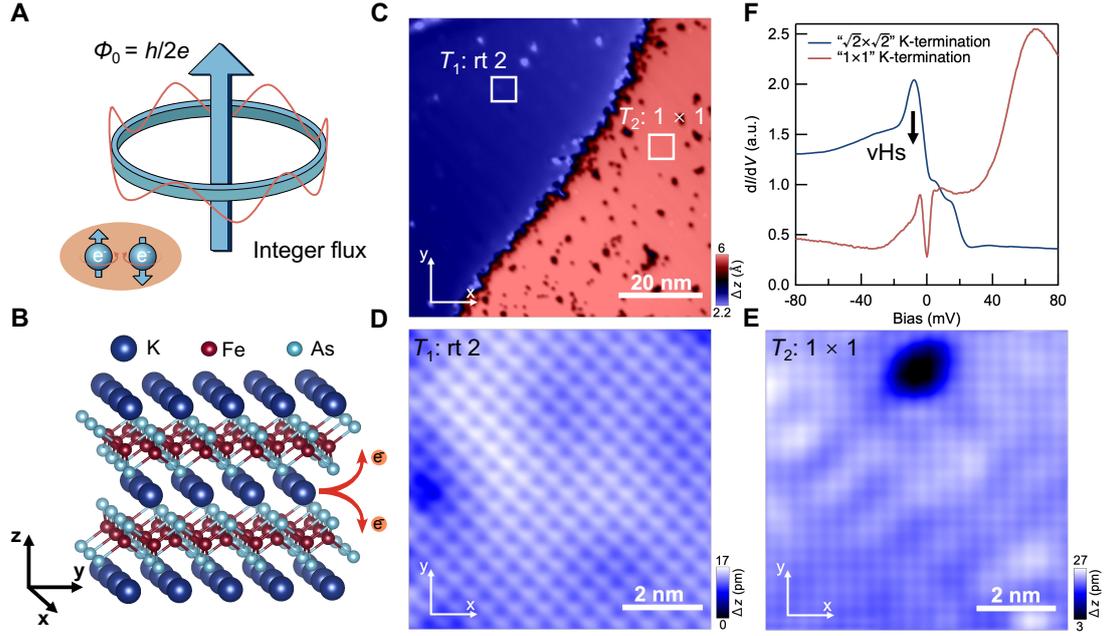

**Fig. 1 Topographic characterization of K-terminated surface.** (**A**) Schematic of magnetic vortices carrying a quantized magnetic flux $\Phi_0 = h/2e$. (**B**) Crystal structure of $KFe_2As_2$ (side view). K atoms act as electron donors, overall providing equal numbers of electrons to the adjacent $Fe_2As_2$ layers. (**C**) STM topographic image of the K-terminated surface. The red and blue regions represent the "1 × 1" and "rt2" ($\sqrt{2} \times \sqrt{2}$) K-terminated surface respectively [$(V_s, I_s)$ = (80 mV, 100 pA), image size: (72 × 72) $nm^2$]. The observed height difference (~ 3 Å) between the two termination types stems from variances in the density of states across their respective. (**D** and **E**) Atomically-resolved topographic images of the (**D**) "rt2" [$(V_s, I_s)$ = (-4 mV, 5 nA), image size: (7.5 × 7.5) $nm^2$] and (**E**) "1 × 1" K-terminated surface [$(V_s, I_s)$ = (-4 mV, 3 nA), image size: (7.6 × 7.6) $nm^2$]. Images in (**C-D**) were obtained at $T$ = 4.2 K, and that in (**E**) at 300 mK. (**F**) Large range $dI/dV$ spectra taken on these two types of K-terminations respectively ["1 × 1": $(V_s, I_s)$ = (80 mV, 1 nA), $V_{mod}$ = 1.25 mV; "rt2": $(V_s, I_s)$ = (80 mV, 500 pA), $V_{mod}$ = 1.25 mV, measured at 4.2 K].

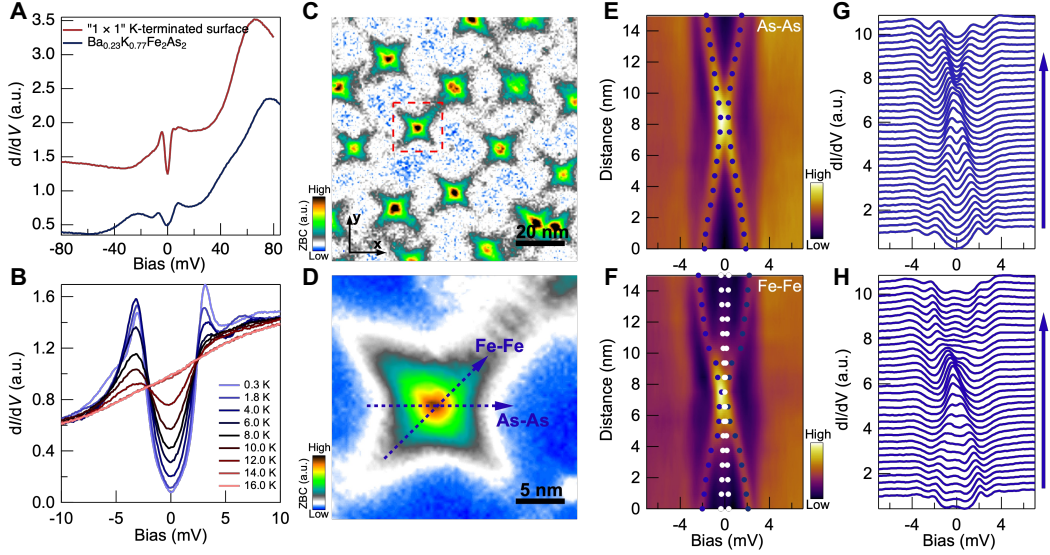

**Fig. 2 Enhanced superconductivity and vortex bound states.** (**A**) Large range d$I$/d$V$ spectra taken at Ba$_{0.23}$K$_{0.77}$Fe$_2$As$_2$ [($V_s$, $I_s$) = (150 mV, 1 nA), $V_{mod}$ = 2 mV] and "1 × 1" K-terminated surface [($V_s$, $I_s$) = (80 mV, 1 nA), $V_{mod}$ = 1.25 mV] measured at 4.2 K. (**B**) Temperature dependent d$I$/d$V$ spectra taken at "1 × 1" K-terminated surface [($V_s$, $I_s$) = (10 mV, 200 pA); $V_{mod}$ = 0.25 mV]. (**C**) Zero-bias conductance (ZBC) map of the "1 × 1" K-terminated surface measured at a temperature of 300 mK with an applied vertical magnetic field ($B_\perp$) of 3T [($V_s$, $I_s$) = (15 mV, 100 pA); $V_{mod}$ = 0.25 mV; image size: (100 × 100) nm$^2$]. (**D**) High resolution ZBC map [($V_s$, $I_s$) = (10 mV, 100 pA); $V_{mod}$ = 0.75 mV; image size: (25 × 25) nm$^2$] taken within the red box in (**C**). The vortex displays a star-shaped morphology, characterized by tails extending along the Fe-Fe axis. (**E** and **F**) Linecut d$I$/d$V$ intensity plots along the blue dashed lines indicated in (**D**). The solid circles indicate the extracted energy positions of the vortex bound states [($V_s$, $I_s$) = (10 mV, 100 pA); $V_{mod}$ = 0.25 mV]. (**G** and **H**) Corresponding waterfall plots of (**E**) and (**F**).

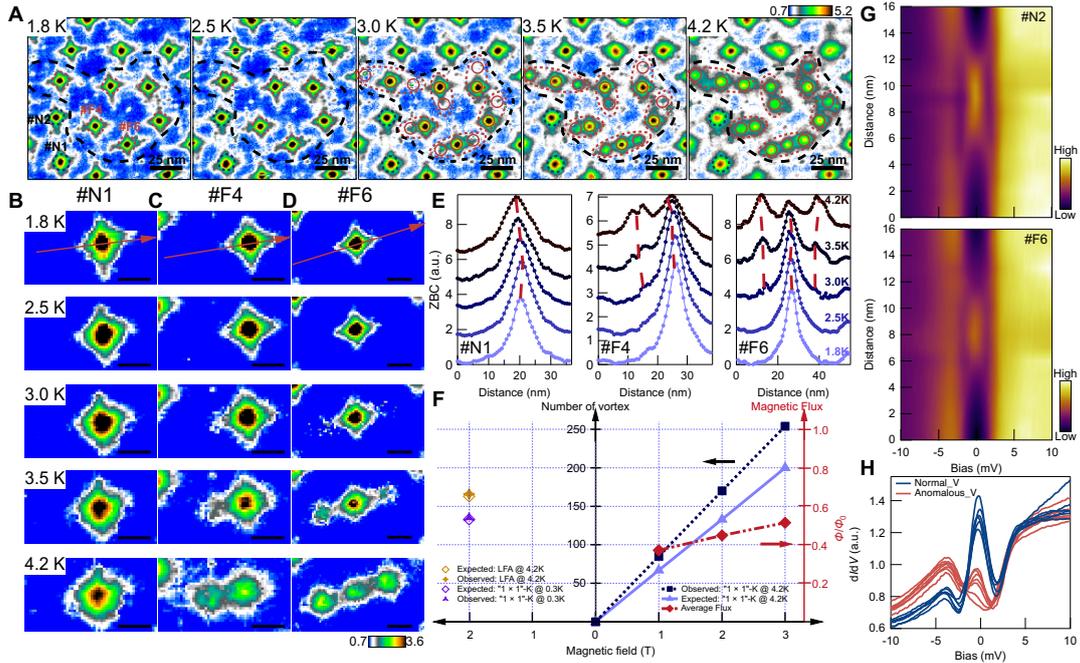

**Fig. 3 Quantum vortex splitting and characteristics of fractional vortices.** (**A**) Temperature dependent ZBC maps recorded with an applied vertical magnetic field of 2 T [($V_s$, $I_s$) = (10 mV, 200 pA); $V_{mod}$ = 0.35 mV; image size: (125 × 125) nm$^2$]. As temperature increases, certain ordinary one-quanta vortices, marked by the red dashed ellipses and solid circles, split into two or three anomalous vortices. (**B** to **D**) Temperature-dependent ZBC maps recorded in the vicinity of vortex cores #N1, #F4, and #F6, as indicated in (**A**). The ZBC intensity at 4.2 K and zero magnetic field was subtracted from the zoomed-in ZBC maps following a detailed topographical correction. The length of the scalar bar is 10 nm. (**E**) Spatial dependence of ZBC along the red arrow lines crossing the vortex cores in (**B** to **D**). As the temperature increases, #N1 (left panel) remains an isolated one-quanta vortex. However, #F4 and #F6 (middle and right panels) progressively split into 2 or 3 anomalous vortices. The red dashed lines indicate the positions of the vortex cores. (**F**) Statistical plot of the number of experimentally observed vortices at different magnetic field conditions, appended with the theoretically expected number of vortices and the average magnetic flux carried by anomalous vortices. (**G**) Line-cut intensity plots across (upper panel) normal vortex #N2 [($V_s$, $I_s$) = (10 mV, 300 pA); $V_{mod}$ = 0.35 mV] and (lower panel) anomalous vortex #F6 [($V_s$, $I_s$) = (10 mV, 200 pA); $V_{mod}$ = 0.35 mV] respectively. (**H**) Point d$I$/d$V$ spectra [($V_s$, $I_s$) = (10 mV, 200 pA); $V_{mod}$ = 0.35 mV] taken from the core centers of multiple normal vortices (blue) and those of anomalous vortices, respectively [($V_s$, $I_s$) = (10 mV, 300 pA); $V_{mod}$ = 0.35 mV; measured at 4.2 K].

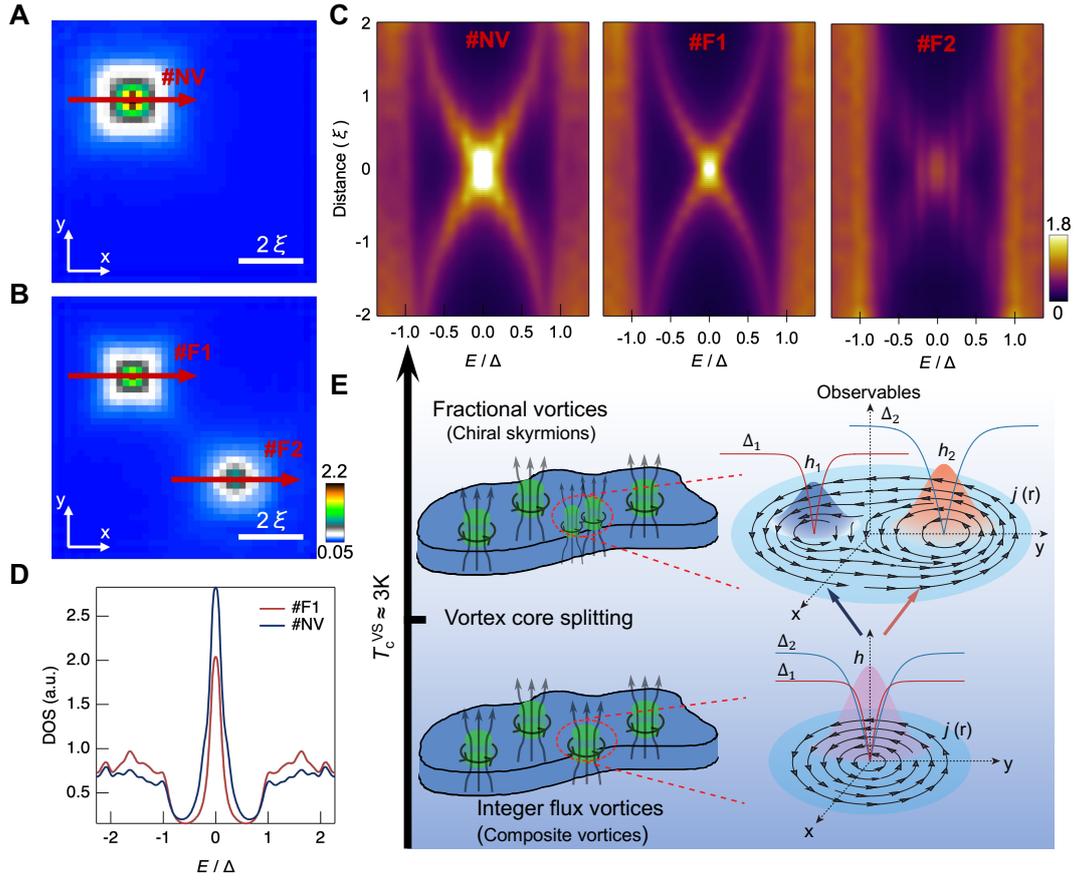

**Fig. 4 Simulation of the vortex splitting and vortex phase diagram.** (**A, B**) Theoretically calculated DOS at zero energy for the normal vortex and the split vortices. (**C**) Calculated line-cut intensity plots across the normal vortex and fractional vortices. (**D**) Calculated DOS spectra at the center of the one-quanta vortex (blue) and the fractional vortex (red). (**E**) Phase diagram of "1 × 1" K-terminated surface. At low temperature, most vortices persist in the form of integer flux vortices (composite vortices). With a gradual increase in temperature ($T_c^{VS}$), thermal fluctuations lead to splitting of some integer flux vortices.

# Supplementary Materials for

# Direct observation of quantum vortex fractionalization in multiband superconductors

Yu Zheng, Quanxin Hu, Haijiao Ji, Igor Timoshuk, Hanxiang Xu, Yongwei Li, Ye Gao, Xin Yu, Rui Wu, Xingye Lu, Vadim Grinenko, Egor Babaev, Noah F. Q. Yuan, Baiqing Lv, Chi-Ming Yim, Hong Ding

Corresponding author: sjtu18810999766@sjtu.edu.cn; baiqing@sjtu.edu.cn; c.m.yim@sjtu.edu.cn

**Materials and Methods**

Scanning tunneling microscopy

The STM/S experiments were performed using a commercial *Unisoku* USM1300 low-temperature STM machine that operates at a base temperature of 300 mK. Pt/Ir tips were used, and conditioned by field emission on a gold target. Differential conductance (d$I$/d$V$) spectra and maps d$I$/d$V$ (**r**, $V$) were recorded using a standard lock-in technique, with the frequency of the bias modulation set at 973 Hz. KFe$_2$As$_2$ single crystal samples were grown by the KAs self-flux method (*1*). Single crystals with sizes up to (2 × 2 × 0.5) mm$^3$ were cleaved mechanically *in situ* at 12K in the ultra-high vacuum condition (base pressure ≈ 2 × 10$^{-10}$mbar), and then immediately transferred to the microscope head which was pre-cooled at temperature of 4.2 K. The microscope head was further cooled temperature of 300 mK via a He$^3$-based single-shot refrigerator. Magnetic field dependent measurements were performed with the applied magnetic field ramped with a small current ramping rate of 50 mA/s. Temperature dependent measurements were performed at zero or constant applied magnetic field. We have extensively scanned each of the KFe$_2$As$_2$ single crystal samples to search for large and clean "1 × 1" K-terminated surfaces necessary for the measurements. Data were acquired at a temperature of 300 mK unless otherwise specified.

Simulation of fractional vortices in Bogoliubov-de Gennes model

We use two models to calculate STM signatures of fractional vortices.
Model I:

According to the Fermi surface of KFe$_2$As$_2$ (three-hole pockets $\alpha, \beta, \gamma$ at the $\Gamma$ point), in the first model, we choose the following basis to describe the low energy band structures around the $\Gamma$ point

$$|1\rangle = (d_{xz} + \mathrm{i}d_{yz})/\sqrt{2}, \qquad |2\rangle = (d_{xz} - \mathrm{i}d_{yz})/\sqrt{2}, \qquad |3\rangle = d_{xy} \qquad (S1)$$

The effective model at the $\Gamma$-point has the point group $C_{4v}$ of KFe$_2$As$_2$ crystal, whose generators include fourfold in-plane rotation $C_{4z}$ and vertical mirror reflection $\sigma_v$. By symmetry analysis, the 3-band spin-degenerate $k \cdot p$ effective model under the basis $(|1\rangle, |2\rangle, |3\rangle)$ is restricted to the following form:

$$H(\boldsymbol{k}) = \begin{pmatrix} -tk^2 & -\mathrm{i}\lambda_1 k_x k_y + \lambda_2(k_y^2 - k_x^2) & \lambda_3 k_- \\ \mathrm{i}\lambda_1 k_x k_y + \lambda_2(k_y^2 - k_x^2) & -tk^2 & -\lambda_3 k_+ \\ \lambda_3^* k_+ & -\lambda_3^* k_- & -t'k^2 \end{pmatrix} \qquad (S2)$$

where $\boldsymbol{k} = (k_x, k_y)$ is electron momentum with $k_\pm \equiv k_x \pm \mathrm{i}k_y$, $t, t'$ are the hopping amplitudes, and $\lambda_1, \lambda_2, \lambda_3$ are the coefficients of the inter-orbital couplings. When the pairing is introduced, on the full Nambu basis

$$\Psi = \left(c_{1,k\uparrow}, c_{2,k\uparrow}, c_{3,k\uparrow}, c_{1,k\downarrow}, c_{2,k\downarrow}, c_{3,k\downarrow}, c^\dagger_{1,-k\downarrow}, c^\dagger_{2,-k\downarrow}, c^\dagger_{3,-k\downarrow}, c^\dagger_{1,-k\uparrow}, c^\dagger_{2,-k\uparrow}, c^\dagger_{3,-k\uparrow}\right)^\mathrm{T},$$

where $c_{i,k\sigma}, c_{i,k\sigma}^\dagger$ are the annihilation and creation operators of electrons in band $i$ with momentum $k$ and spin $\sigma =\uparrow, \downarrow$, respectively, the Bogoliubov-de Gennes (BdG) Hamiltonian of the system is

$$H_{\text{BdG}}(\mathbf{k}) = \begin{pmatrix} H(\mathbf{k})\sigma_0 - \mu & H_\Delta(\mathbf{k})i\sigma_y \\ H_\Delta^\dagger(\mathbf{k})(-i\sigma_y) & -H^*(-\mathbf{k})\sigma_0 + \mu \end{pmatrix}$$

$$= \begin{pmatrix} H(\mathbf{k}) - \mu & 0 & 0 & H_\Delta(\mathbf{k}) \\ 0 & H(\mathbf{k}) - \mu & -H_\Delta(\mathbf{k}) & 0 \\ 0 & -H_\Delta^\dagger(\mathbf{k}) & -H^*(-\mathbf{k}) - \mu & 0 \\ H_\Delta^\dagger(\mathbf{k}) & 0 & 0 & -H^*(-\mathbf{k}) - \mu \end{pmatrix}, \quad \text{(S3)}$$

where $\mu$ is the chemical potential, $\sigma_0$ is the 2 by 2 identity matrix in the spin space, the Pauli matrices $\sigma_{i=x,y,z}$ denote spin, and the pairing term in the orbital space is

$$H_\Delta(\mathbf{k}) = \begin{pmatrix} \Delta_1(k_x^2 - k_y^2) - i\Delta_2 k_x k_y & \Delta_0 & 0 \\ \Delta_0 & \Delta_1(k_x^2 - k_y^2) + i\Delta_2 k_x k_y & 0 \\ 0 & 0 & \Delta_3 \end{pmatrix} \quad \text{(S4)}$$

Here, $\Delta_{j=0,1,2,3}$ are pairing potentials. To describe superconducting vortices, we write BdG Hamiltonian in real space, hence $\mathbf{k} = -i\nabla$ is an operator in the real space.

First, we use the simplest approximation for the vortex core of integer-flux vortex

$$\Delta_j(r, \theta) = \Delta_j e^{i\theta} \tanh\left(\frac{r}{\xi}\right)$$

where $(r, \theta)$ are the polar coordinates in the real space, and $\xi$ is an approximant for a coherence length.

When the Abrikosov vortex is split into two fractional vortices, in the first model we use the following simplest ansatz

$$\Delta_{j=0,1,2}(\mathbf{r}) = \Delta_j e^{i\theta} \tanh\left(\frac{r}{\xi}\right)$$

$$\Delta_3(\mathbf{r}) = \Delta_3 e^{i\varphi} \tanh\left(\frac{|\mathbf{r} - \mathbf{a}|}{\xi}\right)$$

where the vortex core of $\Delta_{j=0,1,2}$ is at the origin, while the vortex core of $\Delta_3$ is at site $\mathbf{r} = \mathbf{a}$. Here $\varphi$ is the polar angle of the vector $\mathbf{r} - \mathbf{a}$.

Then, the density of states (DOS) at energy $E$ can be obtained by the imaginary part of retarded Green function $\rho(E) = \frac{-1}{2\pi} \text{Im } G^+(x, y, E) = \frac{-1}{2\pi} \text{Im} \frac{|\psi_n(x,y)|^2}{E - E_n + i\eta}$, where $E_n$ and $\psi_n(x, y)$ are the eigenvalues and the corresponding eigenfunctions of $H_{\text{BdG}}$.

Figure 4A and 4B show the DOS of one-quanta vortex (**A**) and two split vortices (**B**) at zero energy. Parameters are $(t, t', \lambda_1, \lambda_2, \Delta_0, \Delta_1, \Delta_2, \Delta_3, \xi, \mu)$=(-1, -0.25, 0.35, 0.5, 0.15, 0.03, 0.03, 0.05, 5, -0.6), and the square lattice has 41 × 41 sites. The broadening parameter is $\eta$=0.01. Figure 4C shows the DOS of one-quanta vortex (left) and two split vortices (right) as functions of energy and position. The parameters $(t, t', \lambda_1, \lambda_2, \Delta_0, \Delta_1, \Delta_2, \Delta_3, \xi, \mu)$ are the same as those in Figure 4A and 4B while the broadening parameter is $\eta$=0.01 and the energy step is $\Delta E$ =0.001.

Model II:

In a fully self-consistent solution, the core of a fractional vortex is not axially symmetric and fractional vortices are connected by domain walls. To address that issue, we use the second simple model of three identical Josephson-coupled bands where gaps are calculated self-consistently in order to go beyond an anzats-based solution. In this section, we describe self-consistent solutions of the Bogoliubov de Gennes model. Numerical results are obtained using mean-field Hamiltonian

$$H = -\sum_{\sigma\alpha}\sum_{<ij>} \exp\{iqA_{ij}\} c_{i\sigma\alpha}^\dagger c_{j\sigma\alpha} + \sum_{i\alpha}(\Delta_{i\alpha} c_{\uparrow i\alpha}^\dagger c_{\downarrow i\alpha}^\dagger + H.c.)/2 + \sum_{plaquettes} B_z^2$$

$$\Delta_{i\alpha} = \sum_\beta V_{\alpha\beta} <c_{\uparrow i\beta} c_{\downarrow i\beta}>$$

Coupling to the magnetic field is considered using a discrete version of Maxwell's equation $\nabla \times \nabla \times A = J$, which determines the connection between $A_{ij}$ and $J_{ij}$.

$$J_{ij} = -2q \sum_{\sigma\alpha} Im <c_{i\sigma\alpha}^\dagger c_{j\sigma\alpha} \exp\{iqA_{ij}\}>$$

Tunneling conductance in the system is calculated as

$$\frac{\partial I_{i\alpha}(V)}{\partial V} \propto \sum_n [|u_{in}|^2 f'(E_n - eV) + |v_{in}|^2 f'(E_n + eV)]$$

where $f'$ is a derivative of the Fermi function, $i$ indicates the lattice point of a square lattice, and $n$ denotes the eigenstate of the system. The solution for these equations is obtained using the approximate Chebyshev spectral expansion method, described in (*2*). For further numerical examples and additional detail, see also (*3*).

In this section, we consider a model with three symmetric bands. To minimize the effects of the boundary of the numerical grid, we investigate the case with relatively large intraband interaction $V_{11} = V_{22} = V_{33} = 2.4$, and dimensionless charge $q = 0.25$ to ensure that the vortex solutions are much smaller than the system size. The interband coupling are negative $V_{12} = V_{13} = V_{23} = -0.6$. The simulation results are shown in Fig. S8. For Fig. S8B and Fig. S8C, the temperature is $T = 0.24$ (in bandwidth units); for Fig. S8D and Fig. S8H, it is set to $T = 0.23$. Figure S8D and S8H are performed in an external magnetic field $B = 0.33$. The initial guess for Fig. S8B has $2\pi$ phase winding in all three bands, producing an integer vortex. For Fig. S8C, $2\pi$ winding is imposed only in one of the components, yielding a fractional vortex. Figure S8D and S8H have a domain wall without any vortices as the initial guess for the computation. Pinning centers for Fig. S8**H** were created by an onsite shift of a chemical potential from zero $\mu = 2.0$.

We find that overall spectroscopic signatures are consistent with the experiment and Model I for the individual fractional vortex and individual integer vortex (shown in Fig. S8**A-C**). Namely, compared to normal vortices, in fractional vortices, the intensity of vortex bound state is weaker and superconducting coherence peaks are stronger.

For the case of the presence of a domain wall, we find the results consistent with Ginzburg-Landau modeling (*4*, *5*) with the domain wall being uniformly decorated by fractional vortices in an external field (Fig. S8**D-G**). An integer flux (composite) vortex can be seen as a bound state of several co-centered fractional vortices. In usual cases, the splitting of a normal vortex into two or more fractional vortices is energetically unfavored due to their coupling to the same gauge field and the interband Josephson coupling. However, when there are different domains of the phases of the superconducting order parameters, the splitting of a normal vortex can have lower free energy when the splitting happens along the domain walls instead of other regions. Domains and hence domain walls, can form when there are three or more components of superconducting order parameters coupled by bilinear Josephson terms. Denote the $n$ superconducting order parameters as $\phi_1, \phi_2, ..., \phi_n$, then the generic Josephson coupling up to the leading order can be written as the following Josephson free energy

$$F_J = \frac{1}{2} \sum_{ij} J_{ij} \left( \phi_i^* \phi_j + c.c. \right)$$

where $J_{ij} = J_{ji}$ denotes the Josephson coupling between $i$ and $j$ order parameters. When $n = 2$, $F_J$ forces the fields $\phi_1, \phi_2$ to be either in-phase or out-of-phase depending on the sign of the prefactor, which does not allow degenerate minimal solutions of $\phi_1, \phi_2$. In this case there are no domain walls. The surface electron doping effect in KFe$_2$As$_2$ makes the proximal layers of the surfaces more similar to Ba$_{1-x}$K$_x$Fe$_2$As$_2$ (x $\approx$ 0.77), exhibiting a BTRS superconducting pairing states with three significant Fermi surfaces ($d_{xz}$, $d_{yz}$ and $d_{xy}$ bands) involved in the pairing, as elaborated in above. We thus introduce three order parameters ($n=3$), among them, two order parameters are similar, which we denote as 1 and 2 (formed by $d_{xz}$ and $d_{yz}$ bands). Thus, we can choose $J_{13} = J_{23} \equiv J$ and $J_{12} \equiv J'$, and we can work out the solutions that minimize the Josephson free energy as plotted in Fig. S9. As shown in Fig. S9**A** and **B**, two degenerate minimal solutions can be found as long as $J' > 0$. Given two degenerate solutions, in the thermodynamic equilibrium, time-reversal symmetry would be spontaneously broken, and two types of domains can be formed according to which minimal solution is chosen in the equilibrium state. Domain walls can be left over after a phase transition due to pinning. A domain wall interpolates between two energetically favorable phase-difference-locking values. Then on the domain walls the phase-differences have energetically unfavorable values. If an integer vortex is captured by domain wall, the phase difference values becomes more favorable if it splits into fractional vortices (*4, 5*). Away from domain walls, one-quanta vortices will be formed deep in a given domain. As a result, in the "1 × 1" surface of KFe$_2$As$_2$, both normal vortices with integer flux quantum and anomalous vortices with fractional flux quantum (i.e. fractional vortices) can be observed, and anomalous vortices will distribute along domain walls. Importantly in multiband system there can be more types of vortices than the number of broken symmetries, corresponding to different vorticity in different bands.

Fig. S7 shows the domain wall of the same large field of view at 4.2 K under different magnetic fields along z-axis. However, some of the experimental scans, as shown in Fig. S7C, show that domain walls decorated by fractional vortices "terminate" at two ends. The integer vortices and some pining centers are found around the "termination". To model that situation, we calculated the effects of the presence of pinning centers. In that case, our simulations show that the decoration of domain wall by fractional vortices proceeds only up to the pinning center, where the decoration is terminated by an integer vortex forming on that pinning center (shown on panels Fig. S8**H-K**), for these parameters the spectral features of undecorated-by-vortices domain wall are faint, and should appear as a domain wall "termination" in tunneling experiments, consistently with the experimental observation.

## Supplementary Text

Two types of ordered K-terminated surface

The K-terminated surface shown in Fig S1**A** consists of K atoms in two different surface terminations within the same atomic layer. According to Bardeen model, the tunnelling current can be written as (*6*):

$$I = -\frac{4\pi e}{\hbar} e^{-2z\sqrt{2m\varphi}/\hbar} \rho_t(0) \int_{-eV}^{0} \rho_s(\varepsilon) d\varepsilon \qquad (1)$$

The contrast within a constant current topographic image is not only influenced by the spatial corrugations, but also the integrated local density of states. The topographic image in Fig. S1**A** was recorded with bias $V$ = 80 mV. Due to the higher density of states on the "1 × 1" surface, the distance between the tip and the sample needs to increase to sustain a constant current. This leads to larger apparent height of the "1 × 1" surface (by 3Å) as compared to the "rt2" reconstructed surface, and difference in surface work function between two cleavage planes, i.e. 1.3 eV and 1.6 eV for the "rt2" reconstructed surface and the "1×1" surface respectively (see the Ln(*I*)-Δ*z* spectroscopy data in Fig. S1**C**).

Evidence of electron doping effect on the "1 × 1" surface

As depicted in Fig. 1**B**, the "1 × 1" K-terminated surface contains excess potassium (K) atoms. Utilizing quasiparticle interference (QPI) technique, Fig. S2 shows the interference signals as a function of energy along Γ-M high symmetry direction. Figure S2A shows the schematic fermi surface of KFe$_2$As$_2$. Dark arrows denote the *k*-space scattering process occurring within different pockets, giving rise to the corresponding

$q$-space wave vectors. The electronic band dispersions, energy versus momentum ($E$ - $k$) cuts can then be determined by the measured $q$ geometrically $2|k| = |q|$ along the Γ-M direction. The QPI band dispersion on the "rt2" reconstructed surface primarily originates from surface states (SS) contributed by defects. The energy dispersion of the SS matches well with the measurements (red dashed line) conducted by angle-resolved photoemission spectroscopy (ARPES), confirming the consistency of the doping level between the "rt2" reconstructed surface and the bulk (Fig. S2**B**). On the "1 × 1" surface, the QPI band dispersion (blue dashed line) is mainly attributed to the $d_{xz}$ band. However, compared to ARPES measurements (red dashed line), the QPI band dispersion is shifted approximately 10 meV downwards, indicating a significant electron doping effect on the "1 × 1" surface (Fig. S2**C**).

DFT calculation

Our DFT study is performed by using the full potential local orbital (FPLO) basis(*7*) and the generalized gradient approximation (GGA) to the exchange and correlation functionals (*8*). We use the interpolations of lattice parameters (*9*) and optimize the antimony positions within the GGA. The charge doping of $Ba_{0.23}K_{0.77}Fe_2As_2$ is modeled via the virtual crystal approximation on the Ba/ K site. On the other hand, the band structure of 1-layers $KFe_2As_2$ with "1 × 1" K-terminated surface is estimated by using rigid-band approximation.

Origins of vortex bound states

Through zero-bias conductance mapping we found that the vortex cores observed on the K-terminated surface have a star-like shape with the tails along the Fe-Fe direction. The surface electron doping effect in $KFe_2As_2$ makes the proximal layers of the surfaces more similar to $Ba_{1-x}K_xFe_2As_2$ (x ≈ 0.77). At this point, the momentum distribution of the superconducting gap can be compared to the measurement results of heavily hole-doped $Ba_{0.1}K_{0.9}Fe_2As_2$ (*10*). $d_{xz}$ and $d_{yz}$ orbital have similar gaps around Γ point. The outmost $d_{xy}$ orbital is located closest to the zero gap lines of the $s_\pm$ form order parameter. Superconducting gap on $d_{xy}$ orbital is almost zero, which will not contribute the vortex bound state. The decay length of the core states depends on fermi velocities $v_F$ and gap amplitudes Δ. The rounded-square Fermi surface of outmost $d_{yz}$ orbital has larger fermi velocity $v_F$ along the Fe-Fe direction. Consequently, there are extended tails along the Fe-Fe direction. Meanwhile, $d_{xz}$ ($d_{yz}$) orbital possess a nearly isotropic (anisotropic) superconducting gap and a smaller (larger) fermi energy $E_F$. Hence the $d_{xz}$ ($d_{yz}$) orbital related vortex bound state have a larger (smaller) level spacing and is much more isotropic (anisotropic) (*11*).

Topography correction

Positional correction on the scanned area before performing statistical analysis on the number of magnetic flux vortices under different magnetic fields was required for two reasons. First, the scanning characteristics of the piezoelectric scanner tube employed in our STM equipment were calibrated at 4.2 K. The measured lattice constant of "1 × 1" surface at 4.2 K was 3.8 Å, consistent with the reported value.

However, at 0.3 K the measured value became 4.0 Å, indicating a slight but non-negligible change in the characteristics of the piezoelectric scanner tube with varying temperature, which needs to be considered when performing statistical analysis on the vortex number. Second, each of the ZBC maps reported was recorded with a pixel size of (512 × 512), and required 32 hours to acquire, and thermal drift effect was unavoidable for such long hours of data acquisition at 4.2 K. We took two approaches to mitigate the thermal drift effect. Firstly, before each ZBC map measurement, we waited one to two hours for the system to relax, then found the same surface location and started the measurement. Secondly, using the topographic image at 0.3 K as the control image, we used distinctive surface defects as reference points. After topography correction, nearly every position of defects can be uniquely matched.

Vortex splitting and fractional vortices with magnetic field along different direction

The direction of the magnetic field affects the distribution of fractional vortices. A 2T magnetic field inclined at 45 degrees with respect to the z-axis possesses a 1.4T component along the z-axis, yet the distribution of fractional vortices (Fig. S5**G**) differs from that depicted in Fig. S5**F**. Similarly, a 2T magnetic field inclined at 60 degrees with respect to the z-axis possesses a 1T component along the z-axis, yet the distribution of fractional vortices (Fig. S5**H**) also deviates from that shown in Fig. S5**C**. The distribution of fractional vortices can be easily tuned by tilting of the magnetic field.

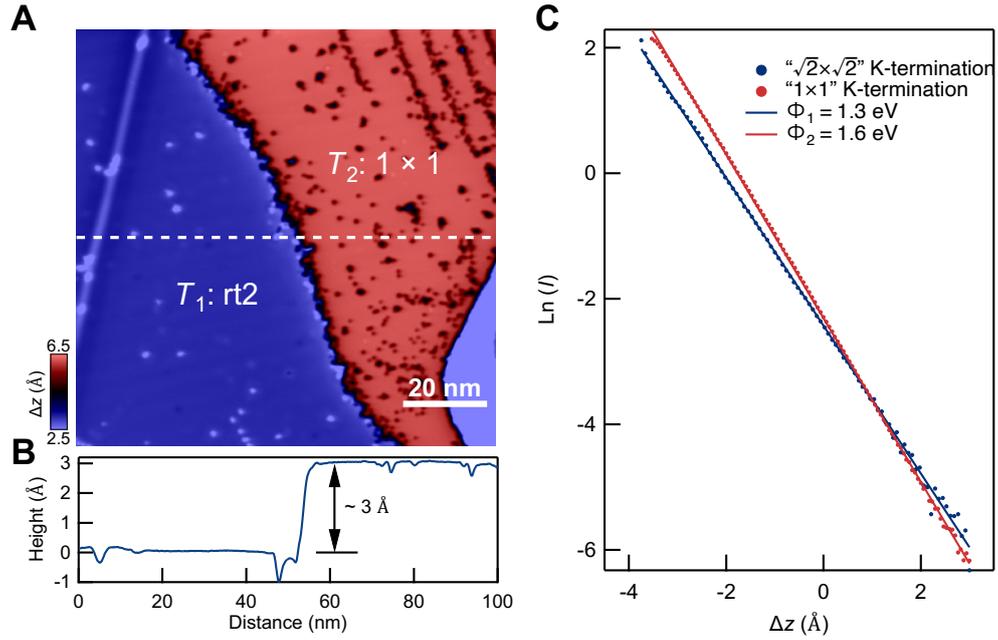

**Fig. S1 Two types of ordered K-terminated surface.** (**A**) Topographic image of a freshly cleaved $KFe_2As_2$ single crystal sample revealing two types of ordered K-terminated surface [$(V_s, I_s)$ = (80 mV, 100 pA), image size: (100 × 100) nm², measured at 4.2 K]. (**B**) Height profile along the white dashed line indicated in (**A**) showing a height difference of ~3 Å between the two types of the K-terminated surface. (**C**) ln($I$)-$\Delta z$ spectroscopy measurements on the two terraces showing the "rt2" terminated surface has a work function of 1.3 eV, while the "1 × 1" surface a work function of 1.6 eV.

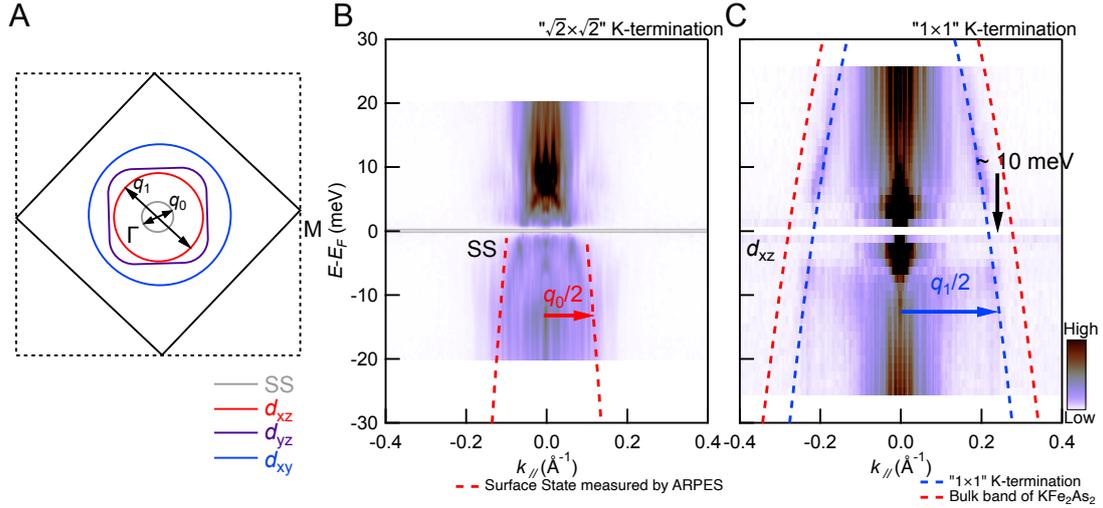

**Fig. S2 Evidence of electron doping effect on the "1 × 1" K-termination. (A)** Schematic fermi surface of $KFe_2As_2$ in the tetragonal nonmagnetic Brillouin zone (solid line). The "one-Fe" zone is shown as a dashed line. The red ($d_{xz}$), purple ($d_{yz}$), and blue curves ($d_{xy}$) show the hole-like pockets surrounding the Γ point. The gray curves show the surface state. Black arrows denote the $k$-space scattering process occurring within different pockets, giving rise to the corresponding $q$-space wave vectors. **(B-C)** STM quasiparticle interference (QPI) energy versus momentum ($E$ - $k$) cuts along the high symmetry Γ-M direction measured on the **(B)** "$\sqrt{2} \times \sqrt{2}$" [($V_s$, $I_s$) = (20 mV, 200 pA); $V_{mod}$ = 0.75 mV; image size: (75 × 75) nm$^2$] and **(C)** "1 × 1" K-terminated surfaces [($V_s$, $I_s$) = (25 mV, 570 pA); $V_{mod}$ = 1.25 mV; image size: (100 × 100) nm$^2$]. In **(B)**, the red dashed line represents the dispersion of the surface state obtained from angle-resolved photoemission spectroscopy (ARPES) measurements. In **(C)**, the blue dashed line represents the extracted energy positions of the holes band measured using STM. The red dashed line represents the dispersion of the $d_{xz}$ band obtained from ARPES. The data was measured at 4.2K.

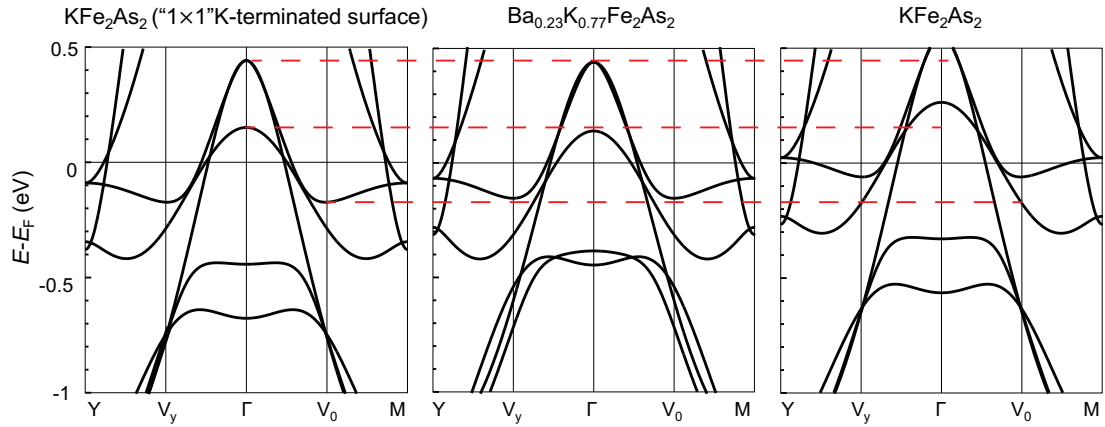

**Fig. S3 DFT calculation.** Calculated band structures of (left) "1 × 1" K-terminated surface of $KFe_2As_2$ (calculated using 1-layer of $KFe_2As_2$ with the "1 × 1" K-terminated surfaces), as well as the bulk band structure of (middle) $Ba_{0.23}K_{0.77}Fe_2As_2$ and (right) $KFe_2As_2$. The red dashed line indicates the band top of the hole band at the Γ point and saddle points in the calculated band structure of the "1 × 1" K-terminated surface. The calculation results show that the resultant doping level of 1-layers $KFe_2As_2$ with "1 × 1" K-terminated surfaces closely resemble that of $Ba_{0.23}K_{0.77}Fe_2As_2$.

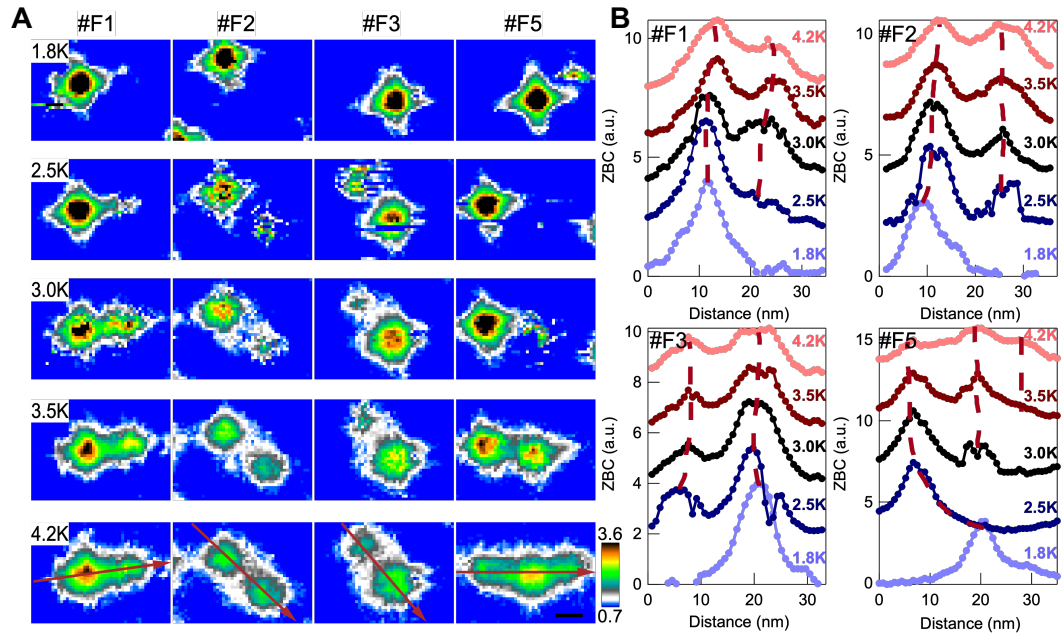

**Fig. S4 Quantum vortex splitting.** (**A**) Temperature-dependent ZBC maps recorded in the vicinity of vortex cores #F1, #F2, #F3 and #F5. The ZBC intensities at 4.2 K and without magnetic field were subtracted from the zoomed-in ZBC maps following a detailed topographical correction. The length of the scalar bar is 7 nm. (**B**) Spatial dependence of ZBC along the red arrowed lines crossing the vortex cores in (**A**).

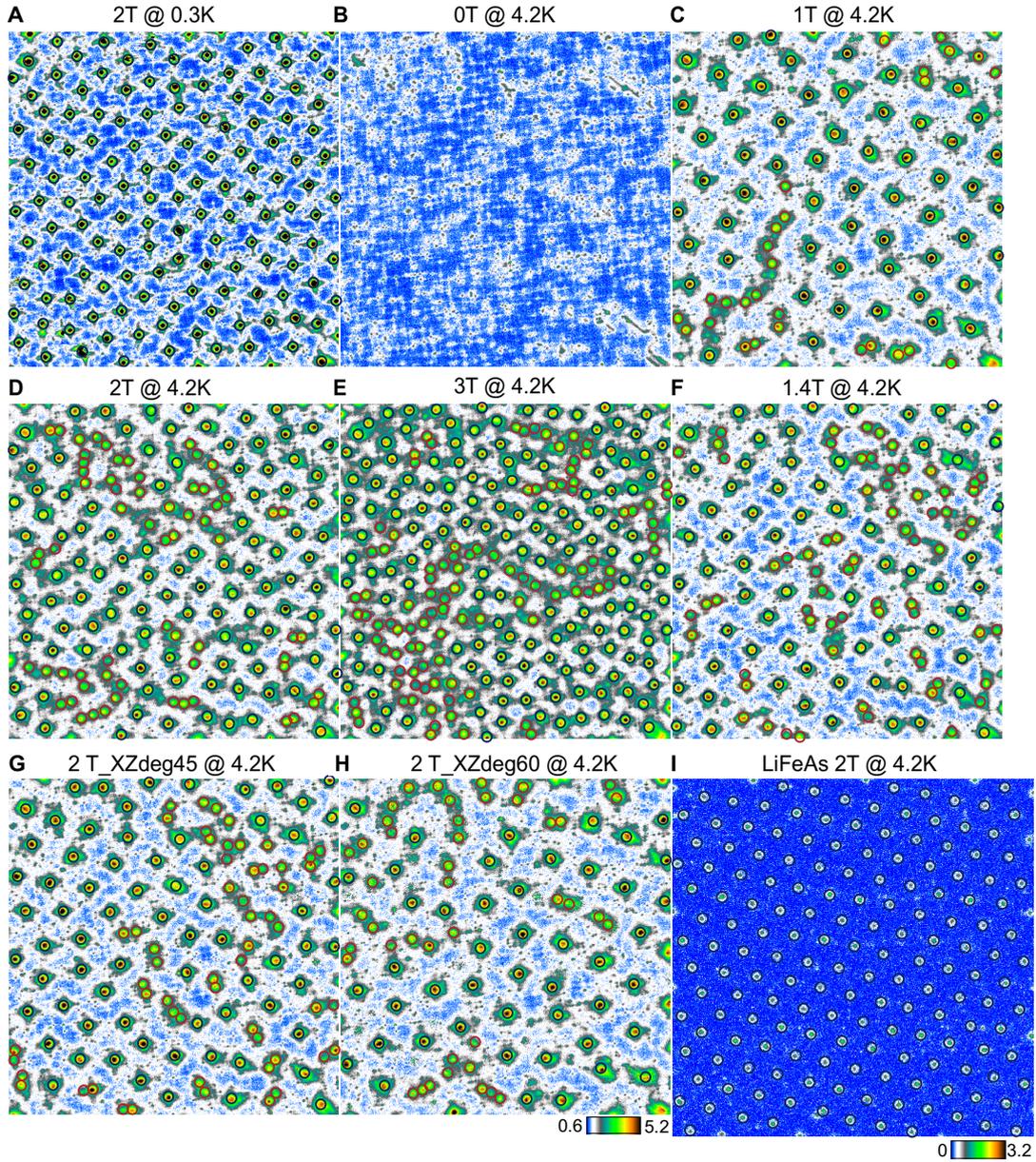

**Fig. S5 Statistical analysis of distinct vortices.** (**A**) Zero-bias conductance (ZBC) map recorded on "1 × 1" K-terminated surface at 0.3 K with an applied magnetic field of 2T, and (**B** to **H**) those recorded in the same field of view at 4.2 K with different applied magnetic fields [$(V_s, I_s)$ = (10 mV, 200 pA); $V_{mod}$ = 0.35 mV; image size: (365 × 375) nm$^2$]. (**I**) ZBC map measured at 4.2K with a magnetic field of 2T on LiFeAs [$(V_s, I_s)$ = (15 mV, 200 pA); $V_{mod}$ = 0.35 mV; image size: (400 × 400) nm$^2$]. The number of vortices present in STM matches what is expected from the number of magnetic flux quanta passing through the area. In (**A**) to (**H**), blue and red circles mark the normal and fractional vortices, respectively. All the ZBC maps shown in (**A**) to (**H**) are recorded at the same surface location following careful topography correction.

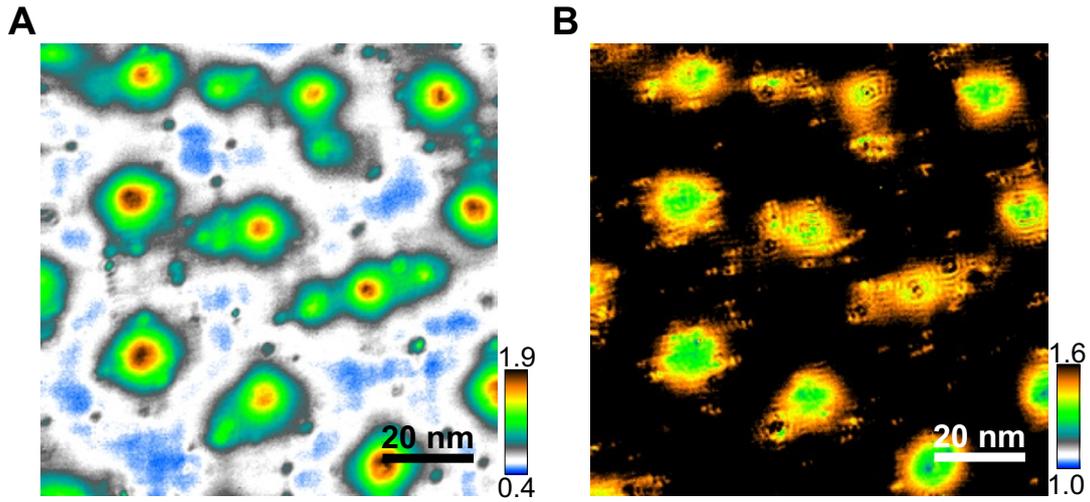

**Fig. S6 Spectroscopy contrast between normal vortices and fractional vortices. (A)** ZBC map and **(B)** d$I$/d$V$(**r**,$V$) map slice at $V$ = -4 mV recorded at the same surface location on the "1×1" K-terminated surface at temperature of 4.2 K and an applied field of 2T [($V_s$, $I_s$) = (10 mV, 200 pA); $V_{mod}$ = 1.25 mV; image size: (100 × 100) nm$^2$]. Compared to normal vortices, fractional vortices have smaller DOS values at 0 meV and larger DOS values at -4 meV.

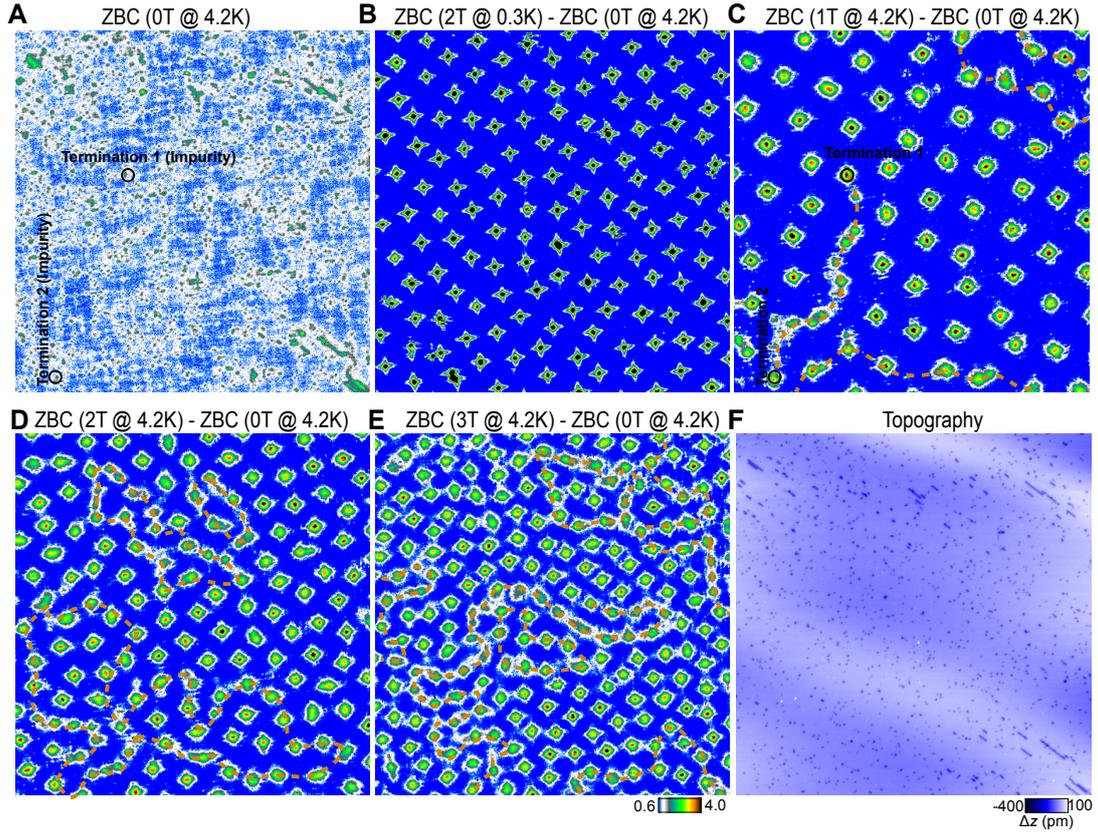

**Fig. S7 Spatial arrangement of the fractional vortices along the expected domain wall.** (**A-E**) Zero-bias conductance (ZBC) maps measured at 4.2 K (**A**) at zero magnetic field and (**B** to **E**) at applied magnetic fields of different field strengths. All the ZBC maps shown in (**B** to **E**) were subtracted from that obtained at zero field in (**A**) following topography correction to eliminate the contributions of the impurities to the image contrast. Distinct linear features that exhibit a higher DOS at zero bias are prominently discernible, and are highlighted using orange dashed line for visual reference. Domain walls decoration by fractional vortices terminates at two ends by pinned integer vortices(**C**). The integer vortices and some pining centers are around the termination. (**F**) Topographic image showing the recorded location of the ZBC maps shown in (**A** to **E**) [($V_s$, $I_s$) = (10 mV, 200 pA); $V_{mod}$ = 0.35 mV; image size: (365 × 375) nm$^2$].

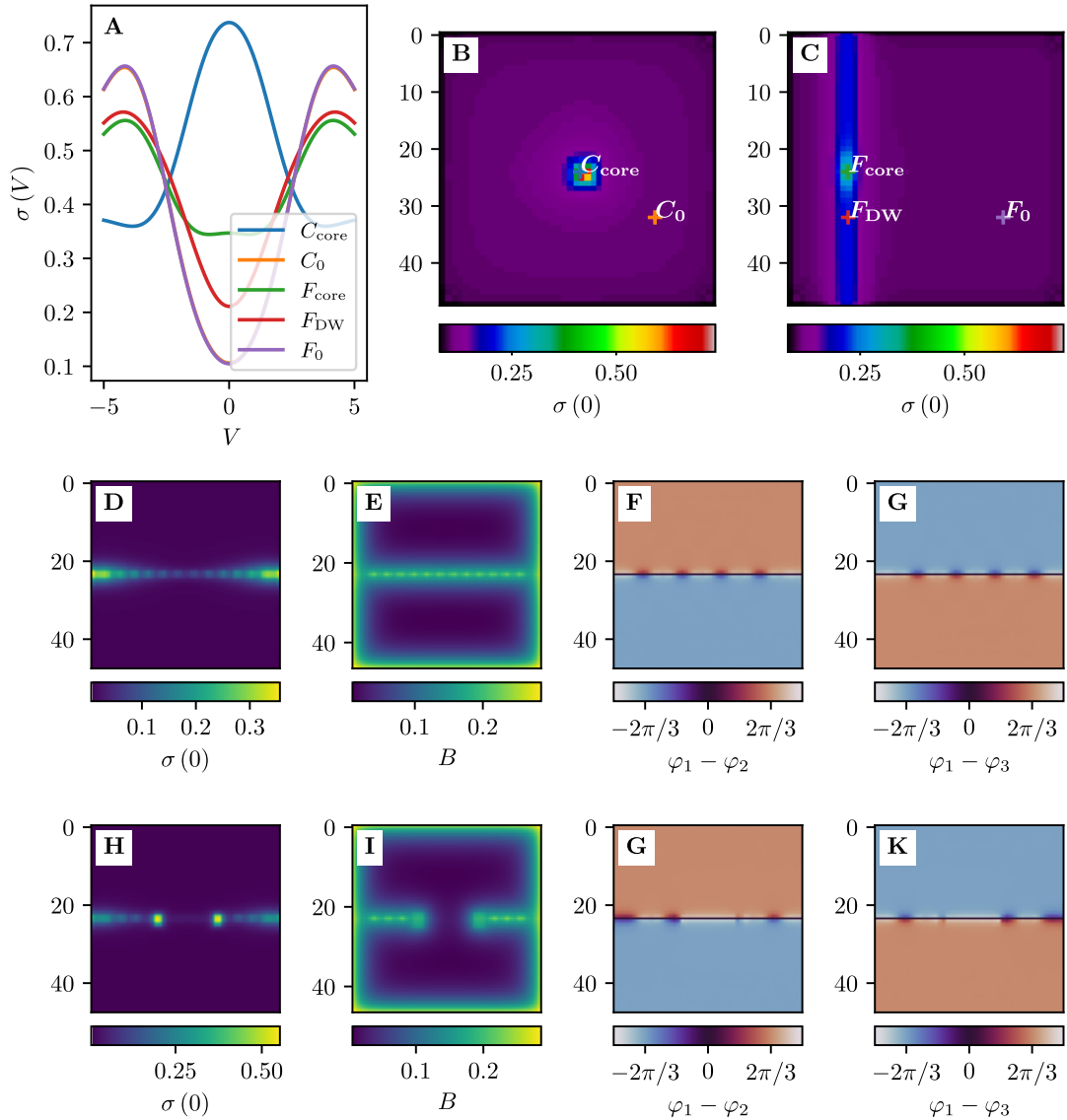

**Fig. S8 Self-consistent BDG numerical results.** (**A**) Tunneling conductance patterns for the conventional vortex core ($C_{core}$), fractional vortex core ($F_{core}$), domain wall ($C_{DW}$), and empty system in both simulations ($C_0$ and $F_0$). (**B**, **C**) Tunneling conductance for zero voltage bias for fractional and conventional vortices. Markers indicate the positions where results for **A** were calculated. (**D**) Tunneling conductance for zero voltage bias for a sample with a domain wall and external magnetic field. Fractional cores are observable. (**E**) Magnetic field distribution in **D**. Vortices cores are easily distinguishable. (**F**, **G**) Phase difference between gaps showing the presence of a domain wall between two relative phase orientations. (**H-K**) The same results with two pinning centers on the domain wall. **H** and **I** show the absence of vortices between the pinning centers; **G** and **K** indicate the presence of the domain wall. Pinning centers stop domain wall decoration by vortices.

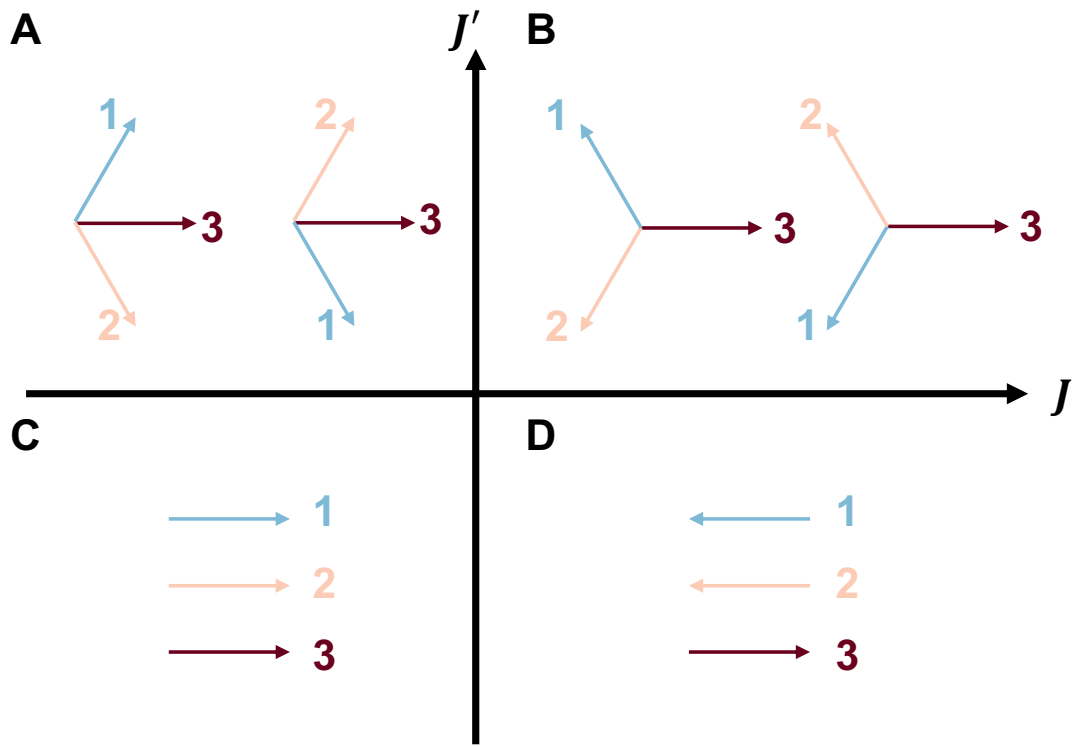

**Fig. S9 Minimal solutions of the Josephson free energy and the phase diagram.** Arrows denote the phases of three order parameters labeled by 1, 2 and 3. **A** and **B** correspond to the case of $J' > 0$ in which there are two degenerate solutions with minimal Josephson free energy, while **C** and **D** to the case of $J' < 0$ where only one solution with minimal Josephson free energy exists. $J < 0$ for **A** and **C**, and $J > 0$ for **B** and **D**.

**Table S1 Number of vortices present in the same STM field of view on the "1 × 1" K- terminated surface at different magnetic field settings and on LiFeAs at vertical field of 2 T ($T$ = 4.2 K)**

| Magnetic field | $N_{theoretical}$ | $N_{experimental}$ | $N_{normal\ vortices}$ ($N_{nv}$) | $N_{anomalous\ vortices}$ ($N_{av}$) | Ratio |
|---|---|---|---|---|---|
| 0 T @ 4.2K | 0 | 0 | 0 | 0 | 0 |
| 1 T @ 4.2K | 67 | 85 | 57 | 28 | 2.80 |
| 2 T @ 4.2K | 133 | 170 | 105 | 65 | 2.32 |
| 3 T @ 4.2K | 200 | 254 | 148 | 106 | 2.02 |
| 2 T @ 0.3K | 133 | 135 | 135 | 0 | 0 |
| 2T_XZdeg60 | 67 | 90 | 47 | 43 | 2.15 |
| 2T_XZdeg45 | 100 | 126 | 70 | 56 | 1.86 |
| 1.4 T @ 4.2K | 100 | 135 | 67 | 68 | 2.06 |
| 2 T @ 4.2K LiFeAs | 163 | 166 | 166 | 0 | 0 |

*Ratio = $N_{anomalous\ vortices}$ / ($N_{theoretical}$ − $N_{normal\ vortices}$)

**References**

1. K. Kihou, T. Saito, K. Fujita, S. Ishida, M. Nakajima, K. Horigane, H. Fukazawa, Y. Kohori, S. Uchida, J. Akimitsu, A. Iyo, C.-H. Lee, H. Eisaki, Single-Crystal Growth of $Ba_{1-x}K_xFe_2As_2$ by KAs Self-Flux Method. *J. Phys. Soc. Jpn.* **85**, 034718 (2016).
2. A. Benfenati, M. Barkman, E. Babaev, Demonstration of CP2 skyrmions in three-band superconductors by self-consistent solutions of a Bogoliubov--de Gennes model. *Phys. Rev. B* **107**, 094503 (2023).
3. I. Timoshuk, E. Babaev, to be published
4. J. Garaud, E. Babaev, Domain Walls and Their Experimental Signatures in s+is Superconductors. *Phys. Rev. Lett.* **112**, 017003 (2014).
5. J. Garaud, J. Carlström, E. Babaev, Topological Solitons in Three-Band Superconductors with Broken Time Reversal Symmetry. *Phys. Rev. Lett.* **107**, 197001 (2011).
6. J. Bardeen, Tunnelling from a Many-Particle Point of View. *Phys. Rev. Lett.* **6**, 57–59 (1961).
7. K. Koepernik, H. Eschrig, Full-potential nonorthogonal local-orbital minimum-basis band-structure scheme. *Phys. Rev. B* **59**, 1743–1757 (1999).
8. J. P. Perdew, K. Burke, M. Ernzerhof, Generalized Gradient Approximation Made Simple. *Phys. Rev. Lett.* **77**, 3865–3868 (1996).
9. H. Chen, Y. Ren, Y. Qiu, W. Bao, R. H. Liu, G. Wu, T. Wu, Y. L. Xie, X. F. Wang, Q. Huang, X. H. Chen, Coexistence of the spin-density wave and superconductivity in $Ba_{1-x}K_xFe_2As_2$. *EPL* **85**, 17006 (2009).
10. N. Xu, P. Richard, X. Shi, A. van Roekeghem, T. Qian, E. Razzoli, E. Rienks, G.-F. Chen, E. Ieki, K. Nakayama, T. Sato, T. Takahashi, M. Shi, H. Ding, Possible nodal superconducting gap and Lifshitz transition in heavily hole-doped $Ba0.1K0.9Fe2As2$. *Phys. Rev. B* **88**, 220508 (2013).
11. N. Hayashi, T. Isoshima, M. Ichioka, K. Machida, Low-Lying Quasiparticle Excitations around a Vortex Core in Quantum Limit. *Phys. Rev. Lett.* **80**, 2921–2924 (1998).